\documentclass[aps,prd,twocolumn,preprintnumbers,superscriptaddress,nofootinbib,amsmath,amssymb]{revtex4}

\usepackage{graphicx}
\usepackage{subfigure}
\usepackage{epstopdf}
\usepackage{dcolumn}
\usepackage{amsmath}
\usepackage[normalem]{ulem}
\usepackage[utf8x]{inputenc}

\usepackage{blindtext}
\usepackage{xcolor}

\newcommand{\beq}{\begin{eqnarray}}
\newcommand{\eeq}{\end{eqnarray}}

\begin{document}

\title{Gauge-invariant screening masses and 
static quark free energies in $N_f = 2+1$ QCD 
at non-zero baryon density.}

\author{Michele Andreoli}
\email{michele.andreoli@pi.infn.it}
\affiliation{Dipartimento di Fisica dell'Universit\`a di Pisa, Largo Pontecorvo 3, I-56127 Pisa, Italy}
\affiliation{INFN - Sezione di Pisa, Largo Pontecorvo 3, I-56127 Pisa, Italy}

\author{Claudio Bonati}
\email{claudio.bonati@df.unipi.it}
\affiliation{Dipartimento di Fisica dell'Universit\`a di Pisa, Largo Pontecorvo 3, I-56127 Pisa, Italy}
\affiliation{INFN - Sezione di Pisa, Largo Pontecorvo 3, I-56127 Pisa, Italy}

\author{Massimo D'Elia}
\email{massimo.delia@unipi.it}
\affiliation{Dipartimento di Fisica dell'Universit\`a di Pisa, Largo Pontecorvo 3, I-56127 Pisa, Italy}
\affiliation{INFN - Sezione di Pisa, Largo Pontecorvo 3, I-56127 Pisa, Italy}

\author{Michele~Mesiti}
\email{michele.mesiti@swansea.ac.uk}
\affiliation{Academy of advanced computing, Swansea University, 
Singleton Park, Swansea SA2 8PP, Wales (UK)}

\author{Francesco Negro}
\email{fnegro@pi.infn.it}
\affiliation{INFN - Sezione di Pisa, Largo Pontecorvo 3, I-56127 Pisa, Italy}

\author{Andrea Rucci}
\email{andrea.rucci@pi.infn.it}
\affiliation{Dipartimento di Fisica dell'Universit\`a di Pisa, Largo Pontecorvo 3, I-56127 Pisa, Italy}
\affiliation{INFN - Sezione di Pisa, Largo Pontecorvo 3, I-56127 Pisa, Italy}

\author{Francesco Sanfilippo}
\email{francesco.sanfilippo@roma3.infn.it}
\affiliation{INFN - Sezione di Roma Tre, Via della Vasca Navale 84, I-00146 Roma, Italy}

\date{\today}

\begin{abstract}
We discuss the extension of gauge-invariant electric and magnetic screening
masses in the Quark-Gluon Plasma to the case of a finite baryon density,
defining them in terms of a matrix of Polyakov loop correlators.  We present
lattice results for $N_f=2+1$ QCD with physical quark masses, 
obtained using the
imaginary chemical potential approach, which indicate that the screening masses
increase as a function of $\mu_B$. A separate analysis is carried out for the
theoretically interesting case $\mu_B/T=3 i \pi$, 
where charge conjugation is not
explicitly broken and the usual definition of the screening masses can be used
for temperatures below the Roberge-Weiss transition. Finally, 
we investigate the
dependence of the static quark free energy on the baryon chemical potential,
showing that it is a decreasing function of $\mu_B$ which displays a peculiar
behavior as the pseudocritical transition temperature at $\mu_B=0$ is
approached.
\end{abstract}

\maketitle

\section{Introduction}
\label{intro}

Static color charges are useful probes of the properties
of strongly interacting matter. 
At low temperature, the 
potential between a heavy quark-antiquark pair, which
can be derived from Wilson loop expectation values 
or from Polyakov loop
correlators, can be used to investigate 
the confining properties of the medium
and the spectrum of heavy quark bound states.
At high temperature, static charge interactions
permit instead to investigate screening effects 
in the Quark-Gluon Plasma (QGP),
which are at the basis of 
interesting phenomenology, like the dissociation
of heavy quark bound states~\cite{matsui} (see
Ref.~\cite{andronic} for a recent review).
Moreover, the asymptotic (large distance) behavior
of Polyakov loop correlators gives access to the 
free energy of static color charges, which is 
a useful probe for confinement/deconfinement.

Interactions between heavy quarks have been widely studied
in lattice QCD simulations by means of Polyakov loop correlators,
in particular by projecting over color group
representations (e.g., singlet or octet) 
after proper gauge fixing. 
Gauge invariant observables can also be studied and, using charge conjugation
symmetry, it is possible to build gauge invariant operators that couple only to
the chromo-magnetic or the chromo-electric 
sector~\cite{Braaten:1994qx,Arnold:1995bh}. In the high
temperature phase correlators of these observables permit to define in a
gauge-invariant and nonperturbative way magnetic and electric screening masses
(inverse of the screening lengths), which have been the subject of recent
lattice QCD investigations~\cite{Maezawa:2010vj,Borsanyi:2015yka}

Screening effects in the QGP are expected to be influenced
by external parameters which can change the properties
of the thermal medium.  In general 
an increase of color screening effects 
(i.e.~an increase of the screening masses) is expected 
as the system is driven deeper into the deconfined region.
An example is the introduction of a magnetic background field
$B$, which 
is known to induce a decrease of the pseudocritical temperature 
$T_c$~\cite{Bali:2011qj} and to affect the confining properties
of QCD~\cite{nostro1, nostro2}, favoring the onset of deconfinement.
Magnetic and electric screening masses in the presence of an external
background field have been determined
by lattice simulations in Ref.~\cite{Bonati:2017uvz}
and have indeed been shown to be  
increasing functions of $B$, in agreement with
analytical studies of screening effects 
in the QGP~\cite{Bandyopadhyay:2016fyd, Hasan:2017fmf, Singh:2017nfa}. 

The baryon chemical potential $\mu_B$ is another parameter
of obvious phenomenological relevance. Also in this case
one expects an increase of screening effects as a function
of $\mu_B$, since a finite baryon density favors the onset of 
deconfinement; this is confirmed by perturbative 
predictions~\cite{lebellac} and by 
lattice QCD studies considering correlators projected
over color representations after gauge fixing~\cite{Takahashi:2013mja}. 
In this case, however, when considering
gauge invariant screening masses
one has to face the problem 
that charge conjugation symmetry is explicitly broken by 
the presence of the baryon chemical potential,
so that a clear separation into electric and magnetic 
sectors cannot be performed any more~\cite{Arnold:1995bh}.

One of the purposes of this study is to propose an extension of the gauge
invariant definition of the screening masses to the case $\mu_B\neq 0$, which
is based on the analysis of a full matrix of Polyakov loop correlators, 
i.e.~including the mixed electric-magnetic correlator
which turns out to be non-zero when $\mu_B \neq 0$. 
Such an extension is then
implemented in numerical simulations of $N_f = 2+1$ QCD
with imaginary values of the baryon chemical potential,
so that the behavior of the gauge invariant 
screening masses as a function of $\mu_B$ is finally
obtained by analytic continuation to real $\mu_B$.
In this way we will show that they
increase as a function of $\mu_B$.
A special attention and a separate 
discussion will be devoted to those  
values of the imaginary chemical potential
for which an exact charge conjugation symmetry
can be recovered (Roberge-Weiss transition points).

A second aim of our study 
is the investigation of the dependence on 
$\mu_B$ of the 
free energy $F_Q$ of a static color charge.
This is related, after proper 
renormalization~\cite{Kaczmarek:2002mc,Petreczky:2004pz,Kaczmarek:2007pb,Kaczmarek:2005gi,Borsanyi:2015yka},
to the large distance behavior of Polyakov 
correlators hence, by cluster property, to the 
Polyakov loop expectation value.
This quantity has been extensively studied in 
the past, at finite $T$ and zero chemical potential,
for its connection to the 
confining properties of the thermal medium.
Here we consider the part of the 
free energy which is related to the introduction 
of the baryon chemical potential,
$\Delta F_Q(T,\mu_B) \equiv 
F_Q(T,\mu_B) - F_Q(T,0)$ 
and can be obtained by studying the ratio of Polyakov loop expectation values.
Also in this case we will present numerical results
obtained for imaginary values of $\mu_B$ and then 
exploit analytic continuation to 
extract the dependence of $\Delta F_Q(T,\mu_B)$
for small values of $\mu_B/T$.

Simulations have been performed on a line of constant
physics of $N_f = 2+1$ QCD with 
physical quark masses, discretized via stout improved staggered fermions.
Most results have been obtained on a $32^3 \times 8$ lattice 
and for temperature above the pseudocritical temperature
$T_c$. In particular, the screening masses have been investigated
for temperatures ranging from $217$ to 
$300$~MeV, while a wider range has been explored
to study the behavior of 
$\Delta F_Q(T,\mu_B)$.
The paper is organized as follows. 
In Section II we review the definition of the gauge invariant screening masses
in terms of Polyakov loop correlators and discuss the extension to the case of
non-zero chemical potential.  We then present the main features of the phase
diagram at imaginary chemical potential and we describe our numerical setup.
In Section~\ref{res} we present our results for the 
screening masses and the quark free energies as a function
of $\mu_B$. Finally, in Section~\ref{concl},
we draw our conclusions.

\section{Observables and Numerical methods}
\label{numsetup}

Our investigation 
is  based on lattice QCD simulations.
As it is well known, the introduction of a baryon chemical
potential $\mu_B$ makes the Euclidean path integral measure
complex, so that standard Monte-Carlo simulations are 
not feasible. A possible solution to this problem, which has been
widely explored in the literature~\cite{alford,lomb99,fp1,dl1,azcoiti,chen,Wu:2006su,NN2011,giudice,ddl07,cea2009,alexandru,cea2012,Karbstein:2006er,cea_other,sanfo1,Takaishi:2010kc,cea_hisq1,corvo,nf2BFEPS,bellwied,gunther,gagliardi,Bornyakov:2017upg}, is 
to perform simulations at imaginary 
values of the chemical potential, $\mu_B = i \mu_{B,I}$,
then exploiting analytic continuation,
which is expected to be valid at least for sufficiently 
small values of $\mu_B/T$.

We will consider a theory with $N_f = 2+1$ flavors (up, down, strange)
and physical quark masses. 
In order to move along a line of non-zero
baryon chemical potential, with electric and strangeness  
chemical potentials set to zero, 
we have chosen degenerate quark chemical potentials
$\mu_{f,I} = \mu_I \equiv \mu_{B,I}/3$. As usual, 
in the discretized
theory, the chemical potentials are introduced in the Dirac operator in 
an exponentiated form attached to temporal gauge 
links~\cite{Hasenfratz:1983ba},
in order to avoid the insurgence of 
ultraviolet divergencies in the continuum limit.
In this way, the introduction of a non-zero $\mu_B$ can also be viewed
as a rotation of temporal boundary conditions for quark fields
by a factor $\exp(\mu_B/(3 T)) = \exp(i \mu_I /T)$.

\subsection{Gauge invariant screening masses at $\mu_B \neq 0$}
\label{obs}

It is well known that a perturbative definition of the screening
masses of QCD~\cite{Gross:1980br,KG_book,Laine_book} gets into
trouble because of non-perturbative contributions 
arising in the finite temperature
gluon propagator at the next-to--leading order,
which are due to magnetostatic gluons and stems from the  
non-Abelian nature of the gauge 
group~\cite{Nadkarni:1986cz,Rebhan:1993az,Rebhan:1994mx,Braaten:1994pk}. 
Nevertheless, it has been shown that well defined screening masses can be
accessed by studying the large distance behaviour of suitable correlators of
gauge invariant quantities, such as the Polyakov loops projected onto the
electric and magnetic sectors~\cite{Braaten:1994qx,Arnold:1995bh}.
In the continuum the Polyakov loop is defined as
\begin{equation}
  \label{eq:polyloop}
  L(\mathbf{r}) =
  \frac{1}{N_c}\mathcal{P}\exp
  \left(ig\int_{0}^{1/T}d\tau A_{0}(\mathbf{r},\tau)\right) ~,
\end{equation}
where $\mathcal{P}$ is the path-ordering operator and $N_c$ is the number of
colors; on the lattice it is easily built in terms of gauge links
in the temporal direction.
 As shown in Ref.~\cite{Arnold:1995bh} (and retraced in some
recent lattice 
works~\cite{Maezawa:2010vj,Borsanyi:2015yka,Bonati:2017uvz}), it is possible
to separate the contributions to screening of the color-magnetic
and color-electric gluons by defining suitable combinations of 
Polyakov loop correlators. 
This is achieved by performing a symmetry decomposition
involving the Euclidean time reversal ($\mathcal{R}$) and the charge
conjugation ($\mathcal{C}$) operators. 
As a result~\cite{Arnold:1995bh, Borsanyi:2015yka,Bonati:2017uvz}, 
one finds that the 
relevant role is played by the fluctuations of the real and imaginary
parts of the Polyakov loop, so that the
correlators
\begin{eqnarray}
  \label{eq:mecorrs}
  C_{M^{+}}(\mathbf{r},T) & = &
  \left\langle \mathrm{TrRe}L(\mathbf{0})\mathrm{TrRe}L(\mathbf{r}) \right\rangle
  - \left\langle \mathrm{TrRe}L \right\rangle^2 ~,\nonumber\\
  C_{E^{-}}(\mathbf{r},T) & = &
  \left\langle \mathrm{TrIm}L(\mathbf{0})\mathrm{TrIm}L(\mathbf{r}) \right\rangle
  - \left\langle \mathrm{TrIm}L \right\rangle^2 ~
\end{eqnarray}
take contributions, respectively, only from the color-magnetic
($\mathcal{R}\text{-even}$ and $\mathcal{C}\text{-even}$) and the
color-electric ($\mathcal{R}\text{-odd}$ and
$\mathcal{C}\text{-odd}$) sectors. In the same way, 
it can be shown~\cite{Borsanyi:2015yka} that the cross sectors
($\mathcal{R}\text{-even}$ and $\mathcal{C}\text{-odd}$ or its
opposite) are trivial: that manifests in the vanishing of the 
mixed correlators between the real and imaginary part of 
the Polyakov loop.
At large distances, these correlators are expected
to behave as~\cite{Braaten:1994qx,Arnold:1995bh} 
\begin{eqnarray}
  \label{eq:mecorrsfit}
  C_{M^{+}}(\mathbf{r},T)\Big|_{r\to\infty} & \simeq & \frac{1}{r}e^{-m_M(T)r} ~,\nonumber\\
  C_{E^{-}}(\mathbf{r},T)\Big|_{r\to\infty} & \simeq & \frac{1}{r}e^{-m_E(T)r} ~,
\end{eqnarray}
where $m_M(T)$ and $m_E(T)$ are the color-magnetic and color-electric
screening masses.

The considerations above are valid when charge conjugation
$\mathcal{C}$ is an exact symmetry of the theory.
The introduction of a baryon chemical potential
$\mu_B$ (or, in general, of chemical potentials
coupled to quark number operators) breaks 
$\mathcal{C}$ explicitly, so that 
a trivial extension of the definition of 
magnetic and electric screening masses 
to finite density QCD is not possible~\cite{Arnold:1995bh}.
What happens is that the electric and 
the magnetic sectors are not separated any more,
a fact that manifests through the appearance
of a non-zero mixed electric-magnetic correlator,
which is defined as follows
(we drop the dependence on $T$ and $\mu_B$ for the sake of
readability):
\begin{equation}
  \label{eq:cxcorr}
  C_{X}(\mathbf{r}) =
  \left\langle \mathrm{TrRe}L(\mathbf{0})\mathrm{TrIm}L(\mathbf{r}) \right\rangle
  - \left\langle \mathrm{TrRe}L \right\rangle \left\langle \mathrm{TrIm}L \right\rangle \, .
\end{equation}
A non-zero value of this correlator means that the real and the 
imaginary part of the Polyakov loop do not undergo independent 
fluctuations any more. The correlator is obviously symmetric
under exchange of electric and magnetic components, 
i.e.~$\left\langle \mathrm{TrRe}L(\mathbf{0})\mathrm{TrIm}L(\mathbf{r}) \right\rangle = 
\left\langle \mathrm{TrIm}L(\mathbf{0})\mathrm{TrRe}L(\mathbf{r}) \right\rangle$
so that, when charge conjugation 
is explicitly broken, 
one can actually define a symmetric matrix of 
correlators
\begin{equation}
  \left(
    \begin{array}{cc}
      C_{M^{+}}(\mathbf{r}) & C_X(\mathbf{r}) \\
      C_X(\mathbf{r}) & C_{E^{-}}(\mathbf{r}) \\
    \end{array}
  \right) ~.
\end{equation}
The practical effect of this mixing is that
the asymptotic, large distance  
behavior of all correlators will now be dominated by a single
mass: that will be confirmed explicitly by our numerical data 
in the following. The information
about the second gauge invariant mass is now hidden
in non-leading corrections to the asymptotic behavior
of the correlators, which are usually difficult to be 
detected directly.

This situation is quite common in the numerical investigation of the 
spectrum of quantum field theories, where one usually considers
a set of mixed correlators in the same channel and needs
to derive the masses of the lightest independent physical states coupled 
to them. A possible solution is to diagonalize the correlator matrix,
i.e.~to solve the eigenvalue equation 
\beq
\mathbf{C} (\mathbf{r})\  \mathbf{x} = C_{1/2} (\mathbf{r})\  \mathbf{x}
\eeq
where $\mathbf{C}$ stands for the matrix and  
the diagonal correlators $C_1$ and $C_2$ are easily found to be 
\begin{eqnarray}
  \label{eq:m1m2corrs}
  C_{1/2}(\mathbf{r})\hspace{-2pt} &=& \hspace{-2pt} 
  \frac{1}{2}\big( C_{M^{+}}(\mathbf{r}) + C_{E^{-}}(\mathbf{r}) \big) \nonumber\\
  & \pm & \hspace{-2pt} 
\frac{1}{2}\Big[ \big( C_{M^{+}}(\mathbf{r}) - C_{E^{-}}(\mathbf{r}) \big)^2 
  + 4 C_{X}^2(\mathbf{r}) \Big]^{\frac{1}{2}} \hspace{-2pt} .
\end{eqnarray}
It is reasonable to expect the long distance behavior 
of the diagonalized correlators to be similar to
that found for the electric and magnetic ones at zero 
$\mu_B$, i.e.
\begin{eqnarray}
  \label{eq:mecorrsfit12}
  C_{1/2}(\mathbf{r})\Big|_{r\to\infty} & \simeq & \frac{1}{r}e^{-m_{1/2} r} ~,
\end{eqnarray}
leading to the definition of two 
independent screening masses
$m_1(T,\mu_B)$ and $m_2(T,\mu_B)$.
Let us discuss the connection between this pair of states with those that can be determined at $\mu_B = 0$:
if at $\mu_B = 0$ the excited state mass $m_M^*$ in the
magnetic sector is higher than $m_E$, these masses satisfy the natural relations
$\lim_{\mu_B \to 0} m_1(T,\mu_B) = m_M(T)$ and $\lim_{\mu_B \to 0}
m_2(T,\mu_B) = m_E(T)$.  In general it is not easy to extract the value of
$m_M^*$ with good precision, however we have verified that in all the cases
explored in this paper our numerical data are consistent with
$m_M^*(\mu_B=0)\ge m_E(\mu_B=0)$.

Let us conclude with some remark on the variational approach used
in the context of hadron 
spectroscopy~\cite{Michael:1985ne,Luscher:1990ck,Blossier:2009kd},
which is different from the one described above. 
In that case, one solves a generalized eigenvalue
problem
\beq
\mathbf{C} (\mathbf{r})\  \mathbf{x} = \lambda_{1/2} (\mathbf{r,r_0})\  
\mathbf{C} (\mathbf{r_0})\
\mathbf{x}
\eeq
where $\mathbf{r_0}$ is a reference distance, which defines the so-called
principal correlators $\lambda_{1/2} (\mathbf{r,r_0})$. We have 
verified that, in the explored cases, the two different approaches
lead to consistent results. The results presented in the
following will be based  
on the diagonal correlators
$C_{1} (\mathbf{r})$ and $C_{2} (\mathbf{r})$.

\subsection{The special case of the Roberge-Weiss point}

As we have already mentioned above, 
the introduction of an imaginary baryon chemical potential
can be rephrased in terms of a modification of the temporal
boundary conditions for all quark fields by a phase 
$\theta~=~\mu_I/T$. That implies that an exact 
charge conjugation symmetry is recovered for special
values of $\theta$: $\theta = \pi$ is one example, however
all values $\theta = (2k + 1)/(3 \pi)$ with $k$ integer 
are equivalent to each other after a global center transformation
on gauge fields~\cite{rwpaper}.
In the case $\theta = \pi$ the charge conjugation
symmetry has the same form as for $\mu_B = 0$,
and it can also be viewed as a switch
from thermal antiperiodic to periodic boundary conditions for 
fermion fields;
 therefore we will take it 
as a reference in the following.

In this case
the standard definition
of electric and magnetic screening masses could be maintained,
however it is well known that for such special values of $\theta$
charge conjugation undergoes a spontaneous breaking above
some critical temperature~\cite{rwpaper}, which
is usually known as the Roberge-Weiss transition temperature
$T_{RW}$ and has been investigated in many
lattice~\cite{finitesize,LPP,FMRW,OPRW,CGFMRW,PP_wilson,wumeng,nagata15,makiyama16,nf2PP,cuteri,noirw}
and model~\cite{model-rw,Sakai:2009dv,sakai2,sakai4,holorw,holorw2,morita,weise,pagura,buballa,kp13,rw-2color} studies.~For $T > T_{RW}$ 
the spontaneous breaking induces a mixed
electric-magnetic correlator, so that one needs to extend the
definition of the gauge invariant screening masses as
discussed above.

On the contrary, for $T < T_{RW}$ one can keep the standard
definition of electric and magnetic sectors 
and ask what modifications they undergo with respect to the standard
case at zero chemical potential. In particular, one expects
a significant change in the behavior of the electric correlator,
since the imaginary part of the Polyakov loop is the order
parameter of the Roberge-Weiss transition at $\theta = \pi$.

\subsection{Dependence of the free energy on $\mu_B$}

The free energy $F_Q$ of heavy quarks in the thermal
medium can be inferred from 
the asymptotic behavior of the unsubtracted Polyakov loop correlator,
i.e.~the squared modulus of $\left\langle \mathrm{Tr}L \right\rangle$,
as follows:
\beq
F_Q = -\frac{T}{2} \log 
|\left\langle \mathrm{Tr}L \right\rangle|^2 \, .~
\eeq

This definition is plagued by additive ultraviolet divergencies
and needs renormalization~\cite{Kaczmarek:2002mc,Petreczky:2004pz,Kaczmarek:2007pb,Kaczmarek:2005gi,Borsanyi:2015yka}, which is usually performed
by subtracting zero temperature contributions.
However the 
introduction of chemical potentials in the discretized 
theory 
is not expected to introduce 
further divergencies, 
at least when this is done 
by exponentiating it in the temporal gauge links~\cite{Hasenfratz:1983ba},  
since that amounts to just
a change of the temporal boundary conditions for fermion
fields, which has no effect at all on the theory in the zero
temperature limit (at least for small enough 
chemical potentials).

For this reason, the contribution to the heavy quark
free energy related to the introduction of a baryon chemical potential,
 \beq
\label{def:deltafq}
\frac{\Delta F_Q(T,\mu_B,\beta)}{T} &\equiv& 
\frac{F_Q(T,\mu_B,\beta) - F_Q(T,0,\beta)}{T} \nonumber \\
&=& -\log \left( 
\frac{|\left\langle \mathrm{Tr}L \right\rangle (T,\mu_B,\beta)|} 
{|\left\langle \mathrm{Tr}L \right\rangle (T,0,\beta)|} 
\right)
\, ,
\eeq
where $\beta$ is the inverse bare gauge coupling,
is expected to be a renormalized quantity with a well 
defined continuum limit. That will be checked explicitly,
based on our numerical results, in Section~\ref{freeene_results}.

\subsection{Simulation details}

We adopted a rooted staggered fermion discretization 
of $N_f=2+1$ QCD. The partition function is
\begin{equation}
  \mathcal{Z} = \int \mathcal{D}U e^{- \mathcal{S}_{\text{YM}}}
  \prod_{f=u,d,s}\det\left[ M_{\text{st}}^{f}(U,\mu_{f,I}) \right]^{1/4} ~, 
\label{partfunc}
\end{equation}
where $ M_{\text{st}}^{f}$ and $\mu_{f,I}$ are respectively 
the fermion matrix and the imaginary chemical potential
for the quark flavor $f$, while
\begin{equation}
  \mathcal{S}_{\text{YM}} = -\frac{\beta}{3} \sum_{i,\mu\neq\nu}
  \left( \frac{5}{6}W_{i;\mu\nu}^{1\times1}
    - \frac{1}{12}W_{i;\mu\nu}^{1\times2} \right)
\end{equation}
is the tree level improved Symanzik gauge action
\cite{Weisz:1982zw,Curci:1983an} and $W_{i;\mu\nu}^{n\times m}$
stands for the trace of the $n\times m$ rectangular parallel transport 
starting from site $i$
and spanning the $\mu$-$\nu$ directions. The
rooted staggered Dirac matrix
\begin{eqnarray}
  M_{\text{st}}^{f}(U,\mu_{f,I}) & = & am_f\delta_{i,j} +
  \sum_{\nu=1}^{4} \frac{\eta_{i;\nu}}{2} \big[
    e^{ia\mu_{f,I}\delta_{\nu,4}}U_{i;\nu}^{(2)}\delta_{i,j-\hat{\nu}} \nonumber
\label{fermatrix}
\\ 
    & - & e^{-ia\mu_{f,I}\delta_{\nu,4}}U_{i-\hat{\nu};\nu}^{(2)\dagger}\delta_{i,j+\hat{\nu}}
        \big]
\end{eqnarray}
is written in terms of two times stout smeared links $U_{i;\nu}^{(2)}$
with isotropic smearing parameter $\rho=0.15$
\cite{Morningstar:2003gk}. Bare parameters entering the action have
been chosen so as to move on a line of constant 
physics~\cite{Aoki:2009sc,Borsanyi:2010cj,Borsanyi:2013bia},
with degenerate light quark masses,
$m_u=m_d=m_l$, 
a physical pion mass $m_{\pi}\simeq135~\text{MeV}$,
and a physical strange-to-light quark mass ratio
ratio $m_s/m_l=28.15$.

Most of our 
Monte-Carlo simulations have been
performed for a fixed value of $N_t$, precisely on a 
$32^3\times8$ lattice. Screening masses have been
measured for four values of the bare
parameter $\beta$ (corresponding to four different lattice
spacings $a$ and temperatures $T=1/(aN_t)$)
and various imaginary chemical potentials $\mu_I$.
The temperature range has been chosen
in order to stay well above $T_{RW}$, which for $N_t = 8$ is 
$T_{RW} \simeq 200$~MeV~\cite{noirw}, so that the
full range of imaginary chemical potentials, up
to the first Roberge-Weiss transition line at $\mu_I/T = \pi/3$, is 
available for analytic continuation.
A summary of the simulation parameters is reported
in Table~\ref{tab:betas}.

No attempt to estimate the magnitude of discretization
effects is performed regarding the screening masses, 
since we have investigated
just one lattice spacing for each temperature;
however we stress that results obtained at $\mu_B = 0$~\cite{Borsanyi:2015yka}
show that, in the same range of temperatures explored in our study,
discretization errors for $N_t = 8$ lattices are not large
and within the statistical accuracy of the results
that we are going to show.

Polyakov loop correlators have been measured, for each run, on a
set of about $5 \times 10^3$ configurations separated by five molecular
dynamics trajectories. As a noise-reduction technique we have
applied to temporal links $3$ steps 
of APE smearing with parameter $\alpha = 0.5$. 
Correlators have been extracted
for generic orientations (i.e.~not just along the lattice axes)
and distances
using a Fast Fourier Transform algorithm, then averaging
correlators corresponding to equal distances
and different orientations.
A blocked jackknife resampling technique has been used in order 
to correctly estimate statistical errors.

In order to extract the screening masses, a fit procedure based on the
model in Eq.~\eqref{eq:mecorrsfit} has been performed,
estimating the statistical errors on the mass parameters by 
resampling techniques. The errors reported in the following will
also include systematic errors estimated by changing the 
range of fitted data points.

As for the determination of the static quark free energy
$\Delta F_Q(T,\mu_B)$, 
given the better statistical accuracy which is reachable 
in this case even with limited statistics, 
we have explored a larger range of temperatures and 
also, in some cases, different values of $N_t$, in order to check 
for finite cutoff corrections. All numerical simulations have been performed
using an RHMC algorithm running on GPUs~\cite{incardona,ferrarapisa}.

\begin{table}[tb]
  \caption{Values of the bare coupling $\beta$ we used for the determination of the screening masses, together with
    the corresponding lattice spacings $a$ 
    and
    temperatures $T$ for our simulations on a $32^3 \times 8$ lattice. 
    In each case we report also the explored
    values of the imaginary quark chemical potential.
   The lattice spacing determination has a systematic uncertainty 
 of the order of $2\text{-}3\%$ \cite{Borsanyi:2010cj,Borsanyi:2013bia}.}  
  \label{tab:betas}
  \begin{ruledtabular}
    \begin{tabular}{ c c c c }
      $\beta$ & $a~[\mathrm{fm}]$ & $T~[\mathrm{MeV}]$ & $\mu_I/(\pi T)$\\
      \hline\\[-1.0em]
      3.94    & 0.0821 & 300 & 0,~0.1,~0.2,~0.3,~1/3\\
      3.8525  & 0.0984 & 251 & 0,~0.1,~0.2,~0.3,~1/3       \\
      3.8225  & 0.1052 & 234 & 0,~0.1,~0.2,~0.3,~1/3       \\
      3.79    & 0.1135 & 217 & 0,~0.1,~0.2,~0.3,~1/3       \\
    \end{tabular}
  \end{ruledtabular}
\end{table}

\begin{figure}[t!]
  \includegraphics*[width=\columnwidth]{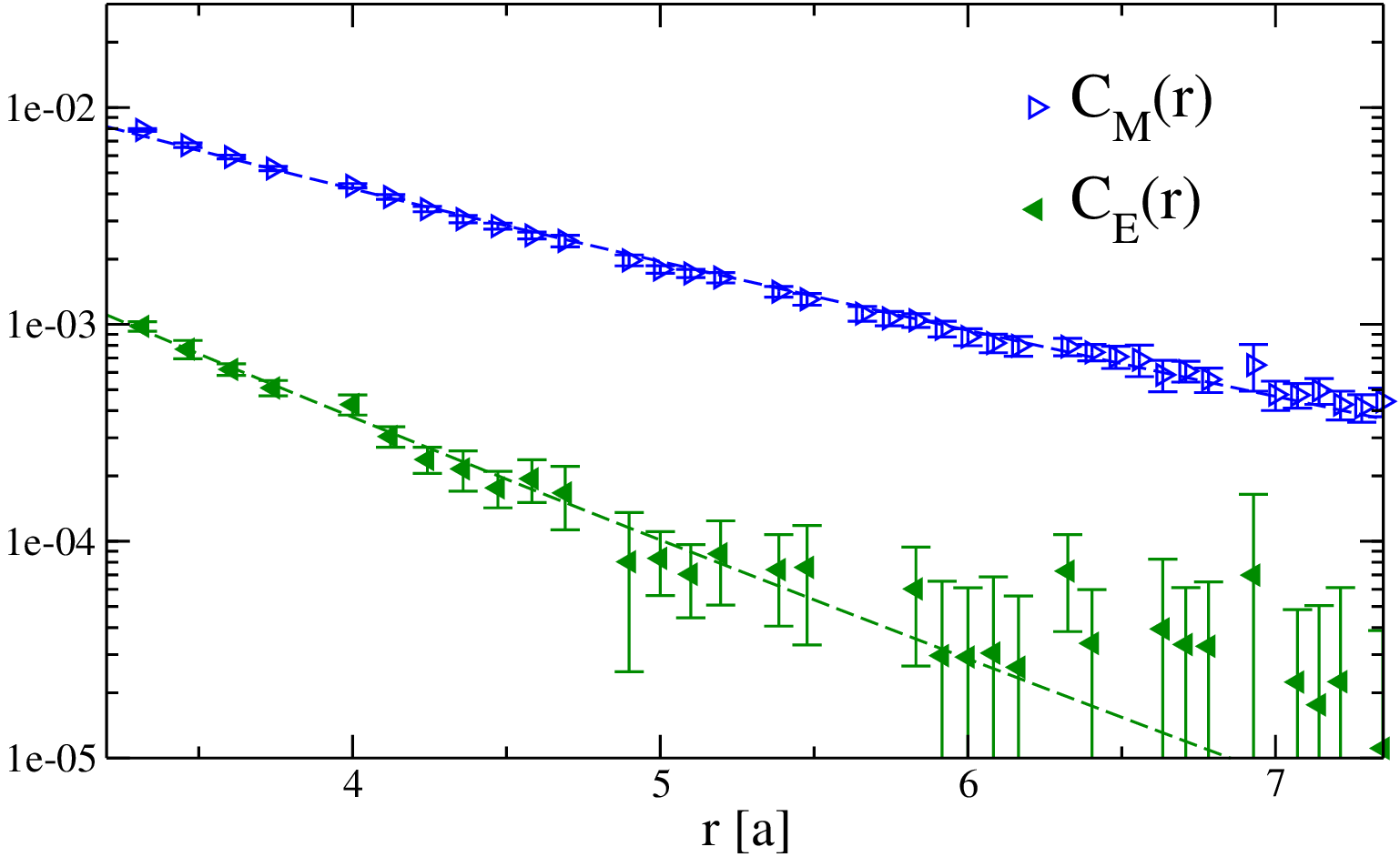}\\~\\
  \includegraphics*[width=\columnwidth]{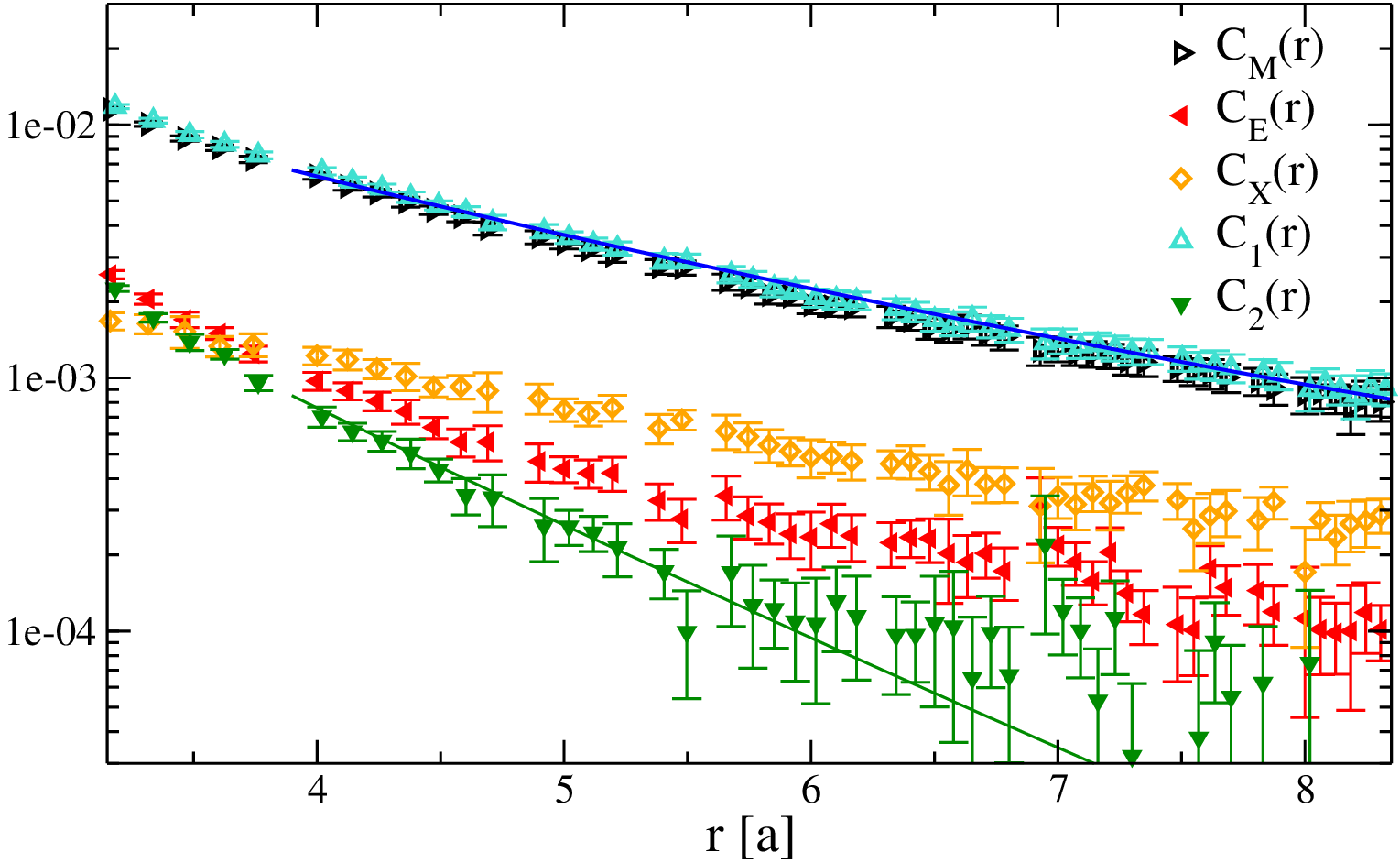}
  \caption{
Behaviour of the color-magnetic $C_M$ and color-electric
    $C_E$ correlators at $T\simeq217~\mathrm{MeV}$ and with $\mu_I/(\pi T)=0$
    (top) and $\mu_I/(\pi T)=1/3$ (bottom). In this latter case, the
    mixed correlator $C_X$ and the 
diagonalized correlators $C_1$ and $C_2$ are also shown for a
    comparison. Best fits to Eqs.~(\ref{eq:mecorrsfit})
and (\ref{eq:mecorrsfit12}) are shown respectively in the two cases.}
  \label{fig:corrs}
\end{figure}

\begin{figure}[t!]
  \includegraphics*[width=\columnwidth]{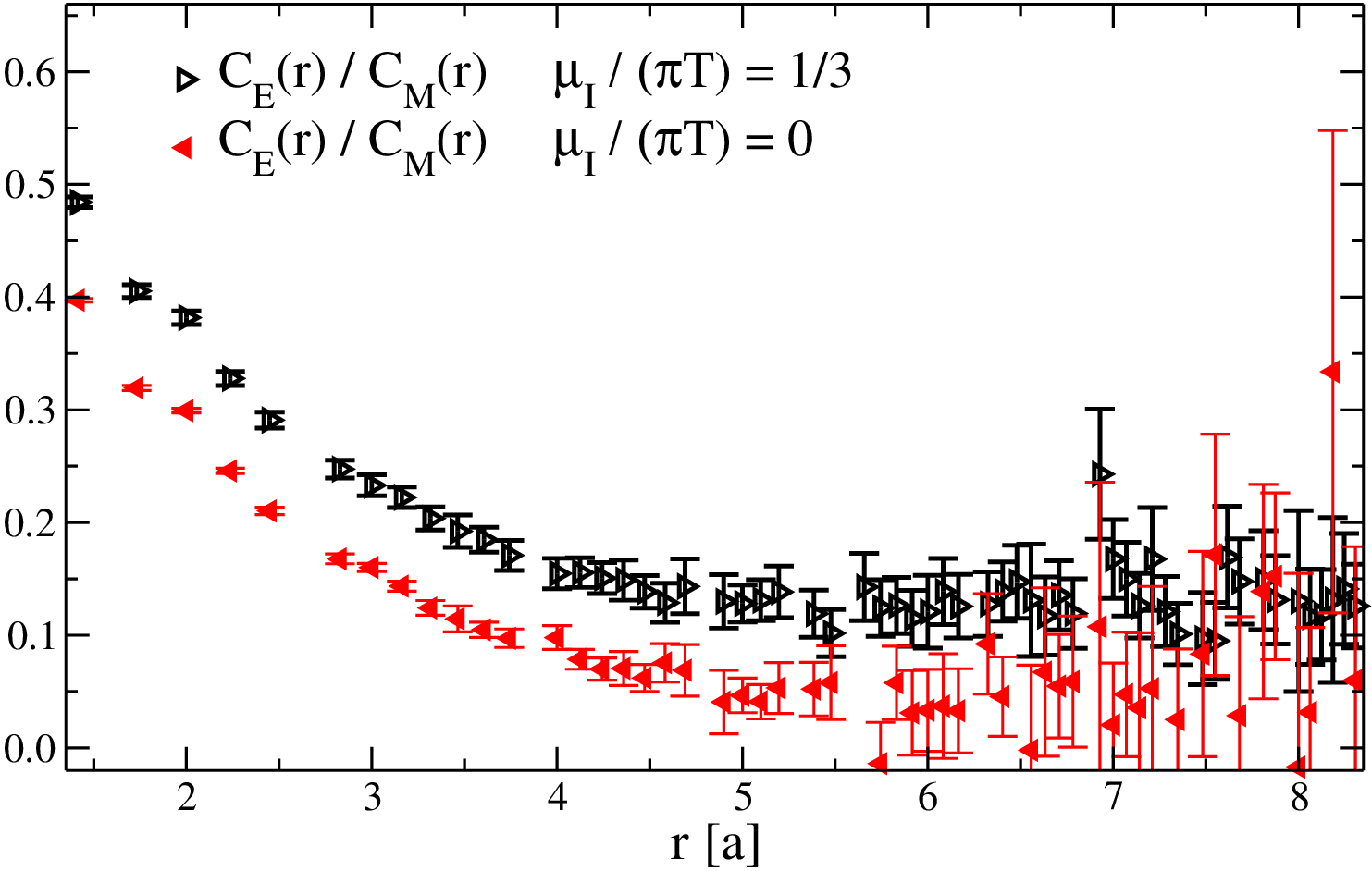}\\~\\
  \caption{Ratios between the electric and the magnetic correlator, as
a function of the distance, for $T\simeq217~\mathrm{MeV}$ 
and respectively $\mu_I/(\pi T)= 1/3$
    and $\mu_I/(\pi T)= 0$.}
  \label{fig:corr_ratios}
\end{figure}

\section{Results}
\label{res}

\subsection{Screening masses}

We start by discussing results obtained for Polyakov loop correlators
and the gauge invariant screening masses. 
In Fig.~\ref{fig:corrs} we report some of the correlators obtained 
for lowest explored temperature, $T~\simeq~217$~MeV.
At zero chemical potential the magnetic and electric correlators,
$C_{M^+}$ and $C_{E^-}$,  show
the standard behavior already clarified in previous 
studies~\cite{Maezawa:2010vj,Borsanyi:2015yka,Bonati:2017uvz}:
the magnetic correlator is larger than the electric
one, by around one
order of magnitude, and decreases 
more slowly as a function of the distance, in agreement with the expected 
hierarchy $m_M < m_E$; best fits according to Eq.~(\ref{eq:mecorrsfit})
are reported together with the data points. The mixed 
electric-magnetic correlator is zero within statistical
errors, as expected, and is not reported in the figure.

In Fig.~\ref{fig:corrs} we report also correlators obtained
at the same $T$ for $\mu_I>0$, in particular for $\mu_I/T = \pi/3$.
In this case the situation is quite different.
The mixed correlator $C_X$ turns out to be different from zero
and, as a result of this mixing, the long-range behavior 
of the electric correlator is modified: indeed, apart from an
overall factor, the magnetic and the electric correlators
show the same behavior at large distances, which seems to be governed 
by the same long distance correlations length.
This phenomenon is more clearly visible in Fig.~\ref{fig:corr_ratios},
where the ratio  $C_{E^-}/C_{M^+}$ is reported as a function
of $r$: while at $\mu_I = 0$ the ratio becomes compatible 
with zero at large distances, at $\mu_I \neq 0$ it approaches
a constant non-zero value, which indicates that the two correlators
fall off in the same way and that their large distance
behavior is dominated by one single mass.

\begin{figure}[t!]
  \includegraphics*[width=\columnwidth]{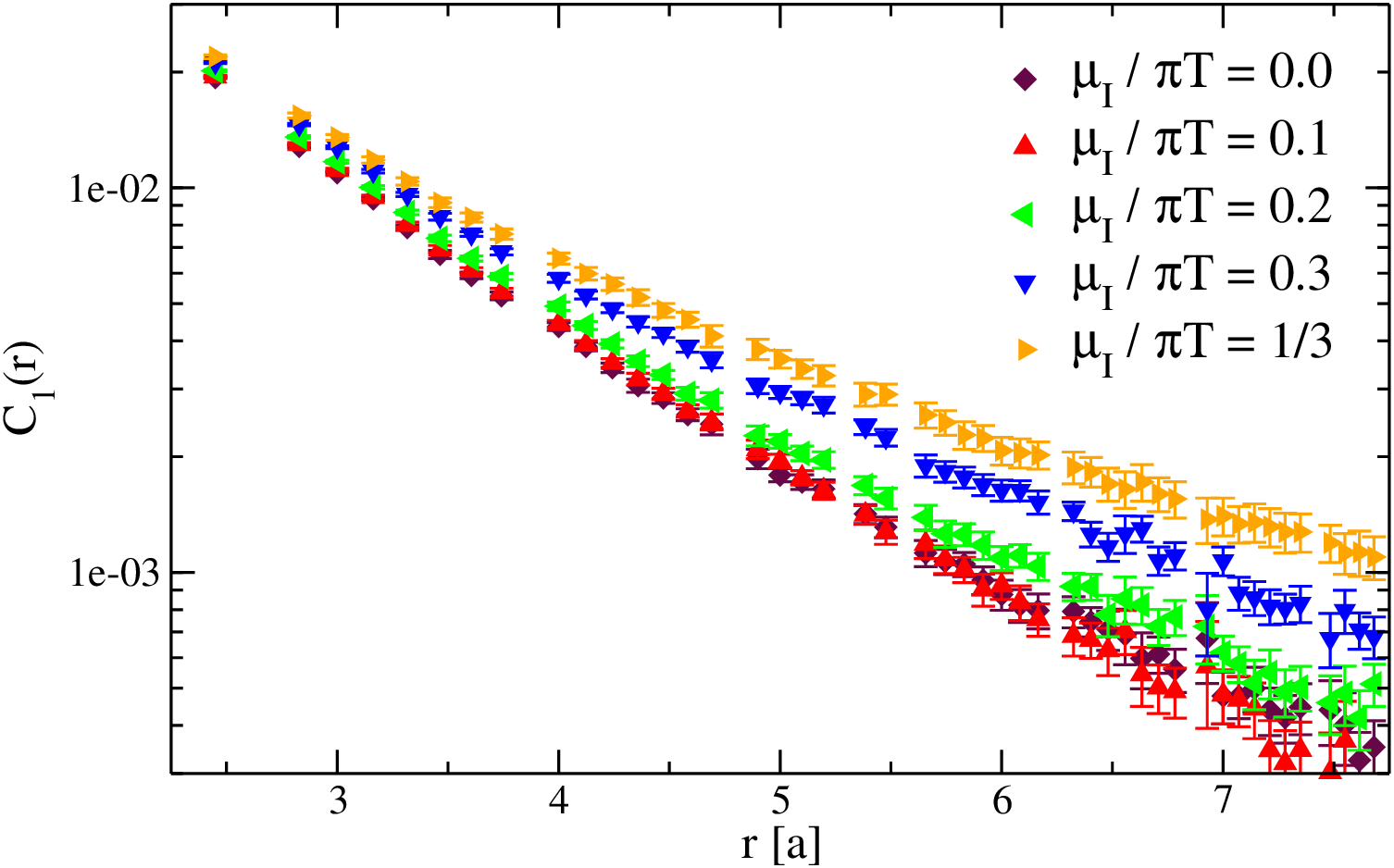}\\~\\
  \includegraphics*[width=\columnwidth]{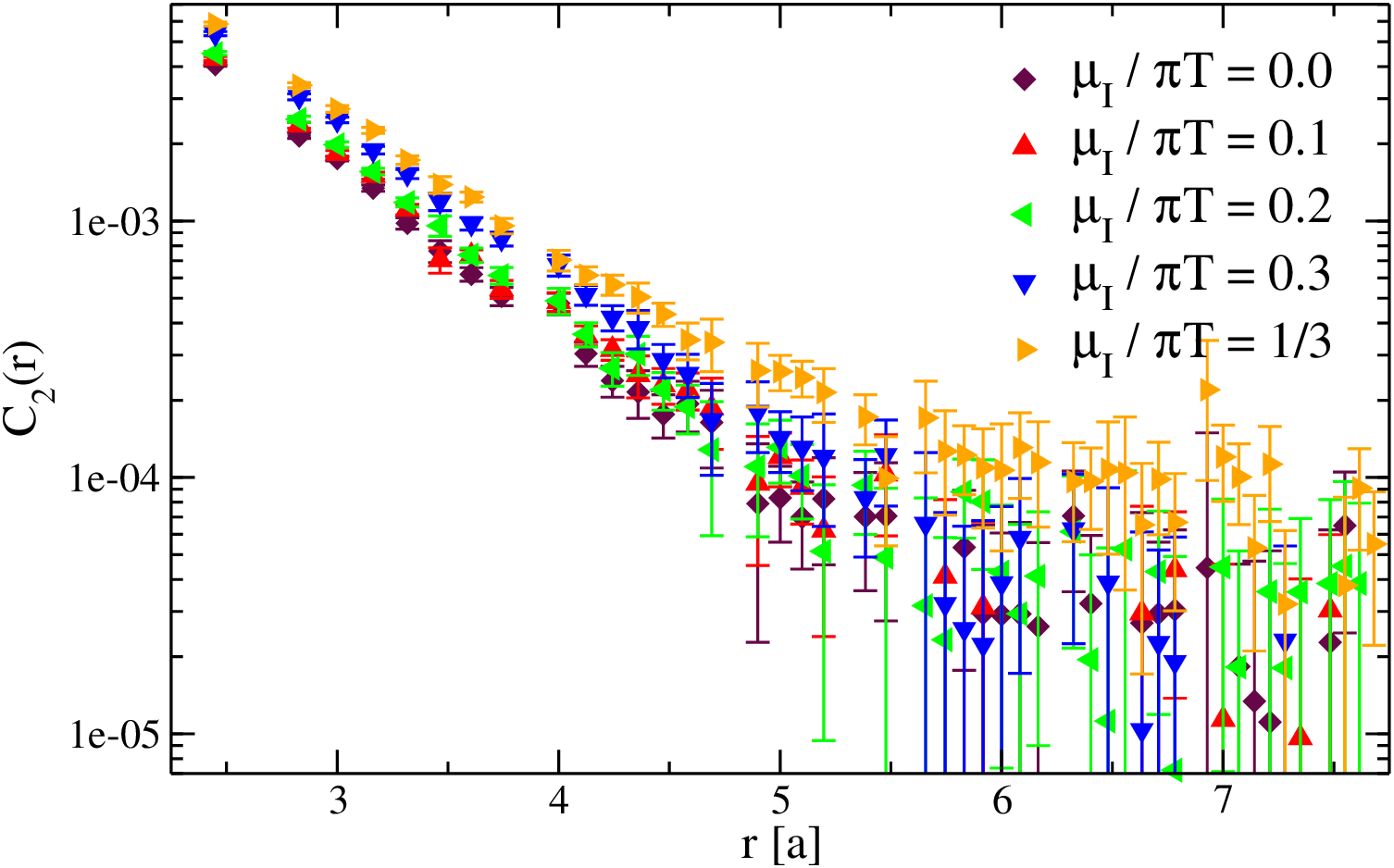}
  \caption{
Behavior of the diagonalized correlators
    $C_1$ and $C_2$ 
at $T\simeq217~\mathrm{MeV}$ for different values of the 
imaginary chemical potential.}
  \label{fig:corrs12}
\end{figure}

This mixing is eliminated when one considers the two diagonalized correlators 
$C_1$ and $C_2$ defined in Eq.~(\ref{eq:m1m2corrs}), which permits
to obtain two well distinct screening masses $m_1$ and $m_2$.
An example of such correlators is shown
in Fig.~\ref{fig:corrs}, together with a best fit according 
to Eq.~(\ref{eq:mecorrsfit12}).
The whole set of diagonalized correlators at $T\simeq 217$~MeV 
is shown in Fig.~\ref{fig:corrs12},
for all the explored values of the imaginary chemical potential.

\begin{figure}[t!]
  \includegraphics*[width=\columnwidth]{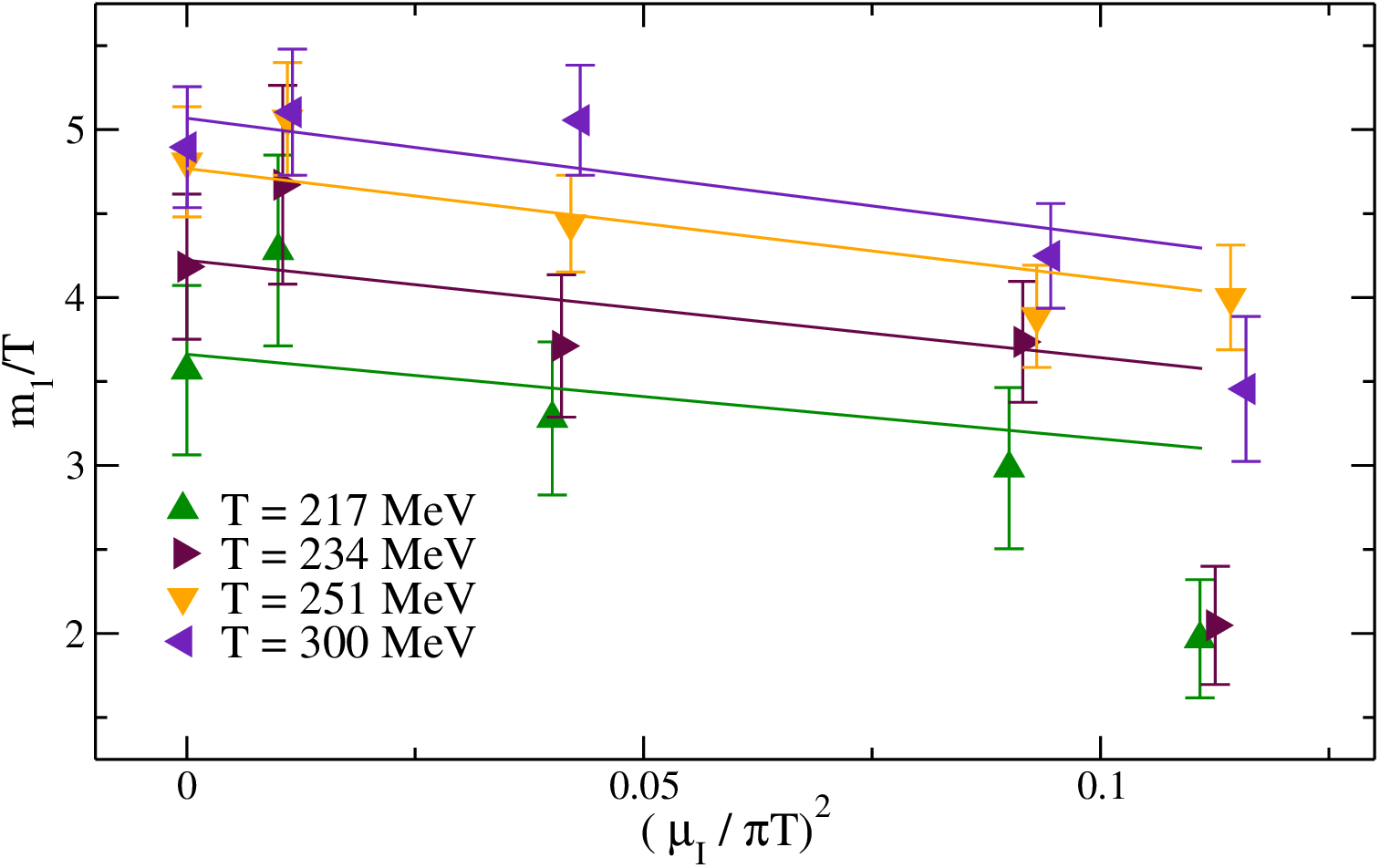}\\~\\
  \includegraphics*[width=\columnwidth]{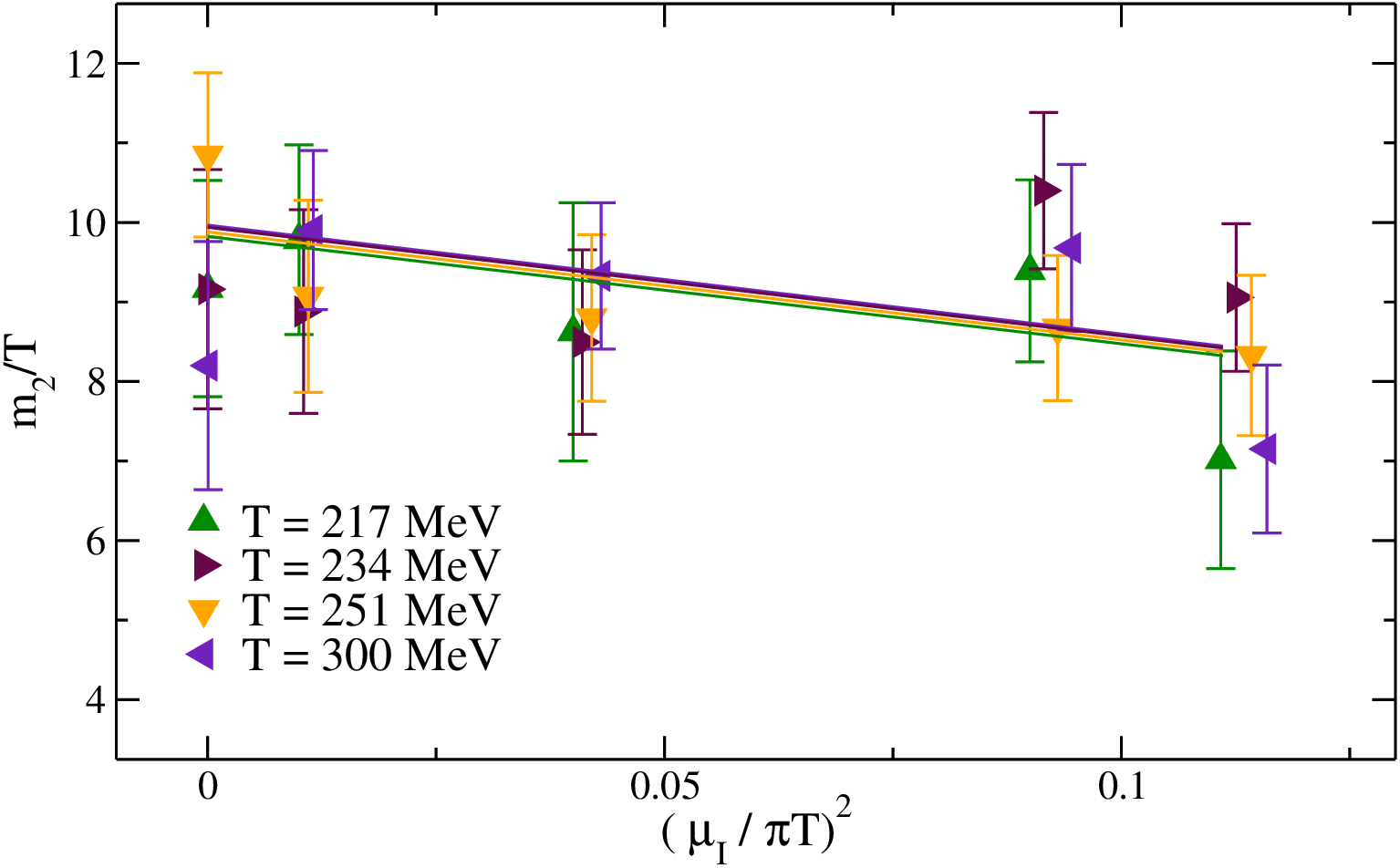}
  \caption{
Diagonalized masses $m_1/T$ and $m_2/T$
as function of $\mu_I/(\pi T)$ for different temperatures.
Curves are the result of a best fit using the ansatz
    in Eq.~(\ref{eq:massfit}) and setting $b_{1/2}(T) = b$ 
where $b$ is a constant (see text).}
  \label{fig:m1m2}
\end{figure}

The gauge invariant screening masses that we have obtained are reported as a 
function of $\mu_I/(\pi T)$ and for various temperatures in Tab.~\ref{tab:m1}
and plotted in Fig.~\ref{fig:m1m2}. Actually, we report the quantity
$m_{1/2}/T$, since this dimensionless ratio is known
to be almost independent of $T$ at zero chemical potential.
Reported error bars include also systematic errors 
related to the choice of fitting range.

The screening masses are expected to be even functions of $\mu_B$. Therefore,
at the leading order in a Taylor expansion, we can write the following
general ansatz:
\begin{equation}
  \label{eq:massfit}
  \frac{m_{1/2}(\mu_B,T)}{T} = a_{1/2} (T) 
\left[1 + b_{1/2} (T) \left(\frac{\mu_B}{3 \pi T}\right)^2\right] \, 
\end{equation}
where $\mu_B = 3 i \mu_I$.
In our fit we have discarded the values obtained for  
$\mu_I/(\pi T) = 1/3$, the reason being that for this value of the 
chemical potential the screening masses are expected
to vanish with a non-analytical behavior 
as the temperature approaches the second order 
Roberge-Weiss endpoint from above~\cite{noirw}.

As a matter of fact, it is possible to fit all data using 
a unique value for the quadratic coefficients,
i.e. setting $b_{1/2}(T) = b$, where $b$ is a $T$ independent
constant, obtaining $b = 1.37(36)$ with 
$\chi^2=15.24/23$. 
A positive value $b$ indicates that screening masses increase
as a function of $\mu_B$:
this is in qualitative
agreement with the fact that the introduction of a 
finite baryon density tends to drive the system farther
away from the confined phase.
The fact that $b$ is independent of $T$ is in agreement with 
leading order perturbative estimates of the 
Debye screening mass~\cite{lebellac} giving
$b = 1.5 N_f/(6 + N_f) = 0.5$
for $N_f = 3$, which is not far from our estimate. The fact
that a common value of $b$ describes both masses implies that
the ratio $m_1/m_2$ is independent of $\mu_B$: notice however that
our data do not put a stringent constraint on this and there
is room for different behaviors within the present errors.

\begin{table}[tb]
  \caption{Screening masses obtained 
for the explored values of $T$ and $\mu_I/(\pi T)$.}
  \label{tab:m1}
  \begin{ruledtabular}
    \begin{tabular}{ c c c c }
      $T$ (MeV) & $\mu_I/\pi T$ & $m_1/T$ & $m_2/T$\\
      \hline\\[-1.0em]
      217 & 0.0 & 3.57(50) & 9.2(1.4) \\
      ''    & 0.1 & 4.28(57) & 9.8(1.2) \\
      ''    & 0.2 & 3.28(46) & 8.6(1.6) \\
      ''    & 0.3 & 2.98(48) & 9.4(1.1) \\
      ''    & 1/3 & 1.97(35) & 7.0(1.4) \\
      \hline
      234 & 0.0 & 4.18(43) & 9.2(1.5) \\ 
      ''    & 0.1 & 4.67(59) & 8.9(1.3) \\
      ''    & 0.2 & 3.71(42) & 8.5(1.2) \\
      ''    & 0.3 & 3.74(36) & 10.4(1.0) \\
      ''    & 1/3 & 2.05(35) & 9.1(0.9) \\ 
      \hline
      251 & 0.0 & 4.81(33) & 10.9(1.0) \\
      ''    & 0.1 & 5.06(34) & 9.1(1.2) \\
      ''    & 0.2 & 4.44(29) & 8.8(1.1) \\
      ''    & 0.3 & 3.89(30) & 8.7(0.9) \\
      ''    & 1/3 & 4.00(31) & 8.3(1.0) \\
      \hline
      300 & 0.0 & 4.90(36) & 8.2(1.6) \\
      ''    & 0.1 & 5.10(38) & 9.9(1.0) \\
      ''    & 0.2 & 5.06(33) & 9.3(0.9) \\
      ''    & 0.3 & 4.25(31) & 9.7(1.1) \\
      ''    & 1/3 & 3.46(43) & 7.2(1.1) \\

    \end{tabular}
  \end{ruledtabular}

\end{table}

\subsection{Results on the dependence of the quark free energy
on $\mu_B$}
\label{freeene_results}

\begin{figure}[t!]
  \includegraphics*[width=\columnwidth]{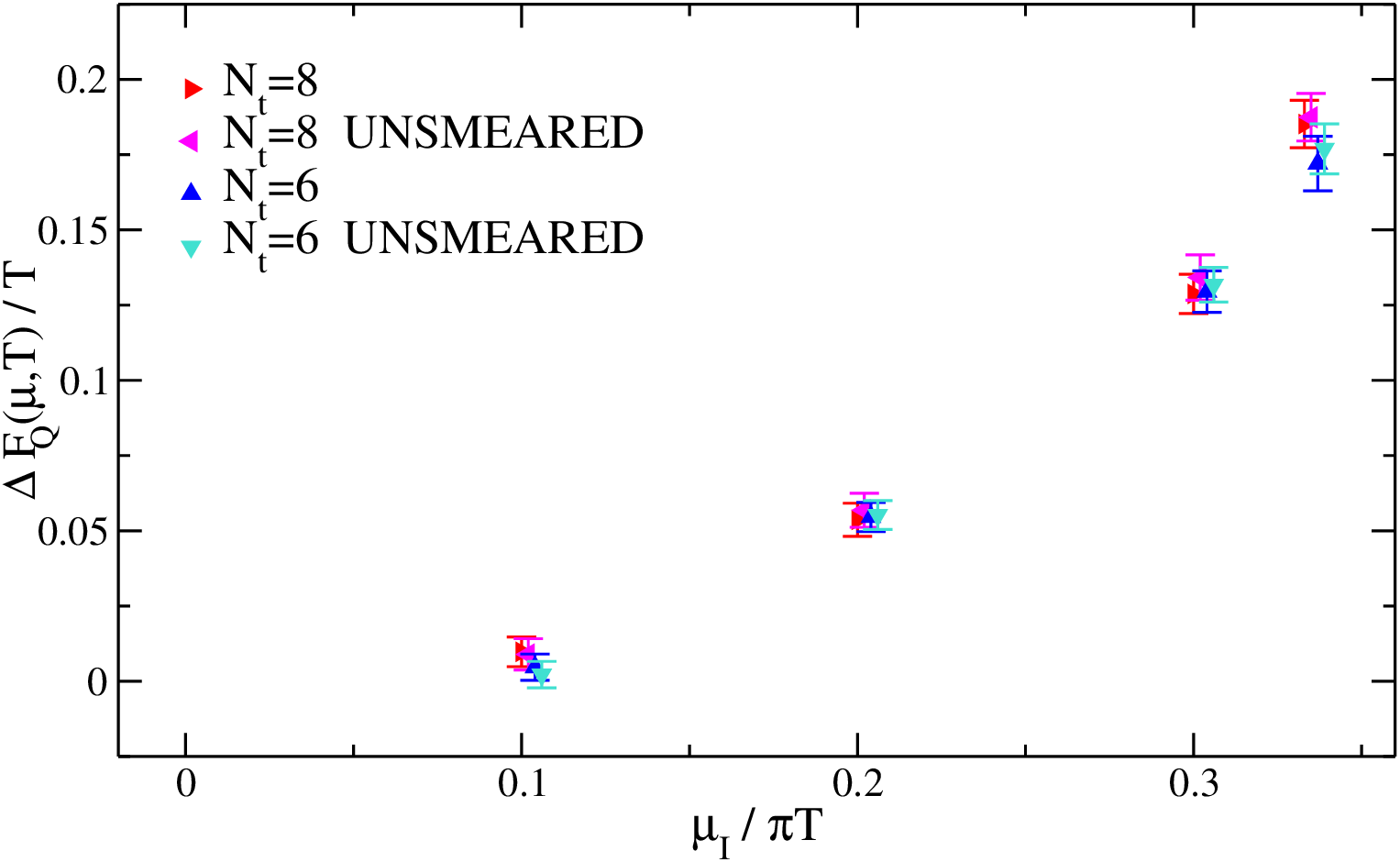}
  \caption{$\Delta F_Q (\mu_B,T)/T$ at $T = 251$~MeV, obtained from the ratio
of Polyakov loops with and without
smearing and for two different values of $N_t$.
}
  \label{fig:freecheck}
\end{figure}

Let us turn to a discussion of the results obtained for 
the dependence of the heavy quark free energy 
on the baryon chemical potential, 
$\Delta F_Q (\mu_B,T)$,
defined
in Eq.~(\ref{def:deltafq}).
To start with, we show in Fig.~\ref{fig:freecheck} the results obtained at 
$T \simeq 251$ MeV for two different values of the temporal 
extension,
$N_t = 6, 8$, and from both smeared and unsmeared
Polyakov loops. The fact that all determinations agree within
errors is a convincing test that this quantity is well
defined and does not need renormalization,
as expected. Moreover, data suggest that $O(a^2)$ corrections
are not significant, within present errors, already for $N_t = 6$.

\begin{figure}[t!]
  \includegraphics*[width=\columnwidth]{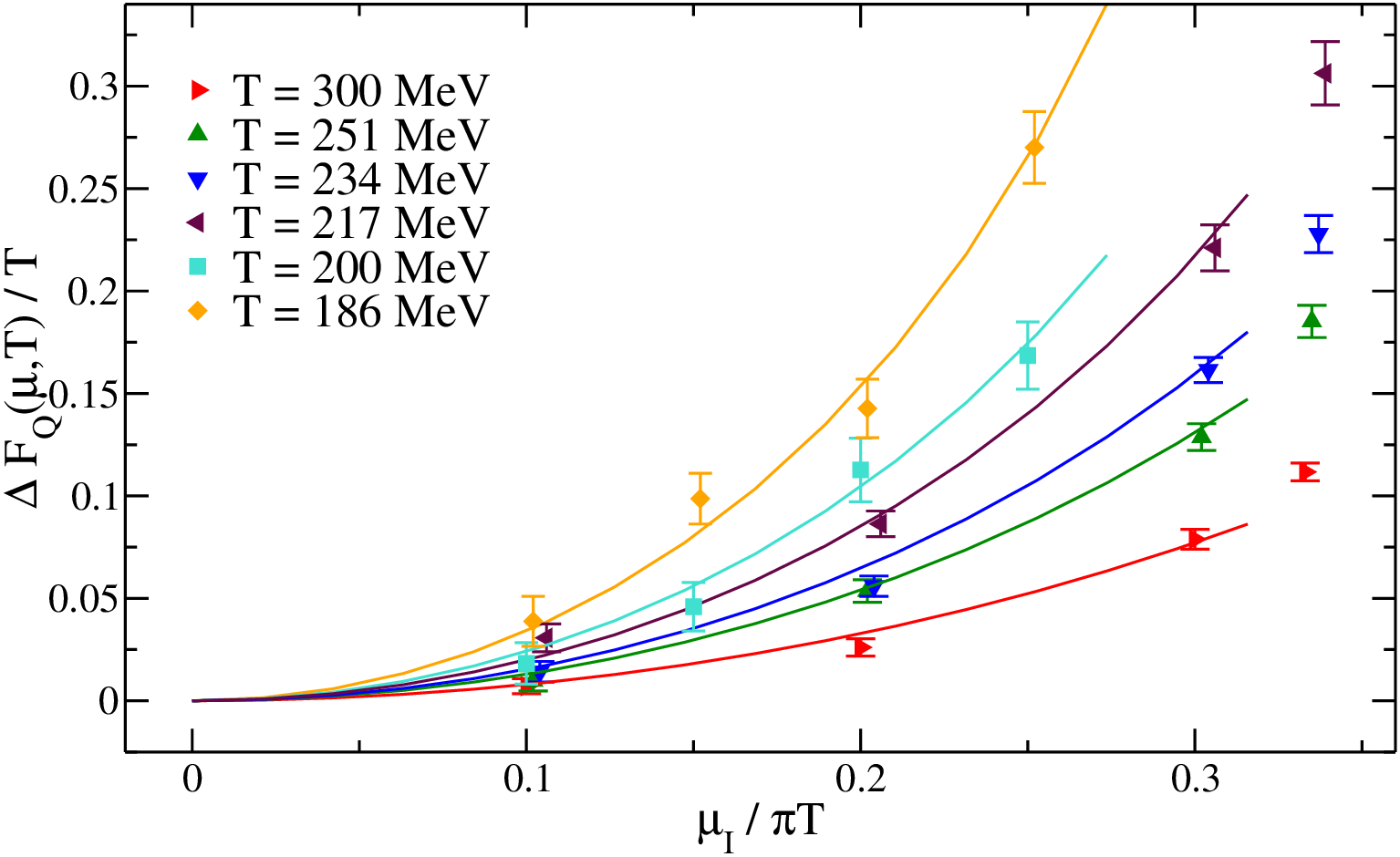}
  \caption{$\Delta F_Q (\mu_B,T)/T$ at various temperatures
obtained from the ratios of smeared Polyakov loops on
lattices with $N_t = 8$.}
  \label{fig:fovert}
\end{figure}

It is not the aim of this study to provide a continuum
extrapolation for $\Delta F_Q (\mu_B,T)$, therefore for the
other temperatures we just provide results obtained for
$N_t = 8$, which are shown for some 
of the explored temperatures in Fig.~\ref{fig:fovert}.
The free energy is expected to be an even 
and analytic function of 
the chemical potential, therefore fitting numerical
data 
with a truncated Taylor expansion 
in $(\mu_B/T)^2$ is the natural choice. Within the statistical 
accuracy of our data we are able to reliably estimate
just the first term in the expansion, moreover we have
noticed that the quality of the fit improves
if one tries to fit directly the behavior of the 
squared ratio of Polyakov loops, i.e.~the exponential
of the free energy, according to 
\beq
\frac{|\left\langle \mathrm{Tr}L \right\rangle (T,\mu_B)|^2} 
{|\left\langle \mathrm{Tr}L \right\rangle (T,0)|^2} &=&  
\exp\left( -2 \frac{\Delta F_Q (\mu_B,T)}{T} \right) 
\label{ratioloopfit}
\\
&=& 1 - \chi_{Q,\mu_B^2} \left(\frac{\mu_B}{T}\right)^2 
+ O \left( \left({\mu_B}/{T}\right)^4 \right) \nonumber
\, ,  
\eeq
where 
\beq
\chi_{Q,\mu_B^2} \equiv \frac{\partial^2 (F_Q/T)}{\partial (\mu_B/T)^2} \, .
\eeq
Such a functional dependence, with higher order terms neglected, 
describes reasonably well data with  $\mu_I/T~<~\pi/3$ 
at all temperatures:
the results of our best fits 
are reported in Table~\ref{tab:freeenergyfit}.

\begin{table}[t!]
  \caption{Results for the coefficient $\chi_{Q,\mu_B^2}$ entering 
the parametrization of $\Delta F_Q (\mu_B,T)$ defined in 
Eq.~(\ref{ratioloopfit}), as obtained from a fit to our numerical
data for $\mu_I/T < \pi/3$.}
  \label{tab:freeenergyfit}
  \begin{ruledtabular}
    \begin{tabular}{ c c c }
      $T$ & $\chi_{Q,\mu_B^2}$ & $\chi^2/\mathrm{dof}$ \\
[+0.4em]
      \hline \\[-1.0em]
      185  & 0.0742(33) & 3.52/3 \\
      200  & 0.0530(32) & 1.64/3 \\
      217  & 0.0440(17) & 2.78/2 \\
      234  & 0.0341(10)  & 3.96/2 \\
      251  & 0.0288(11) & 0.56/2 \\
      300  & 0.0179(9)  & 3.68/2 \\
    \end{tabular}
  \end{ruledtabular}
\end{table}

\begin{figure}[t!]
  \includegraphics*[width=\columnwidth]{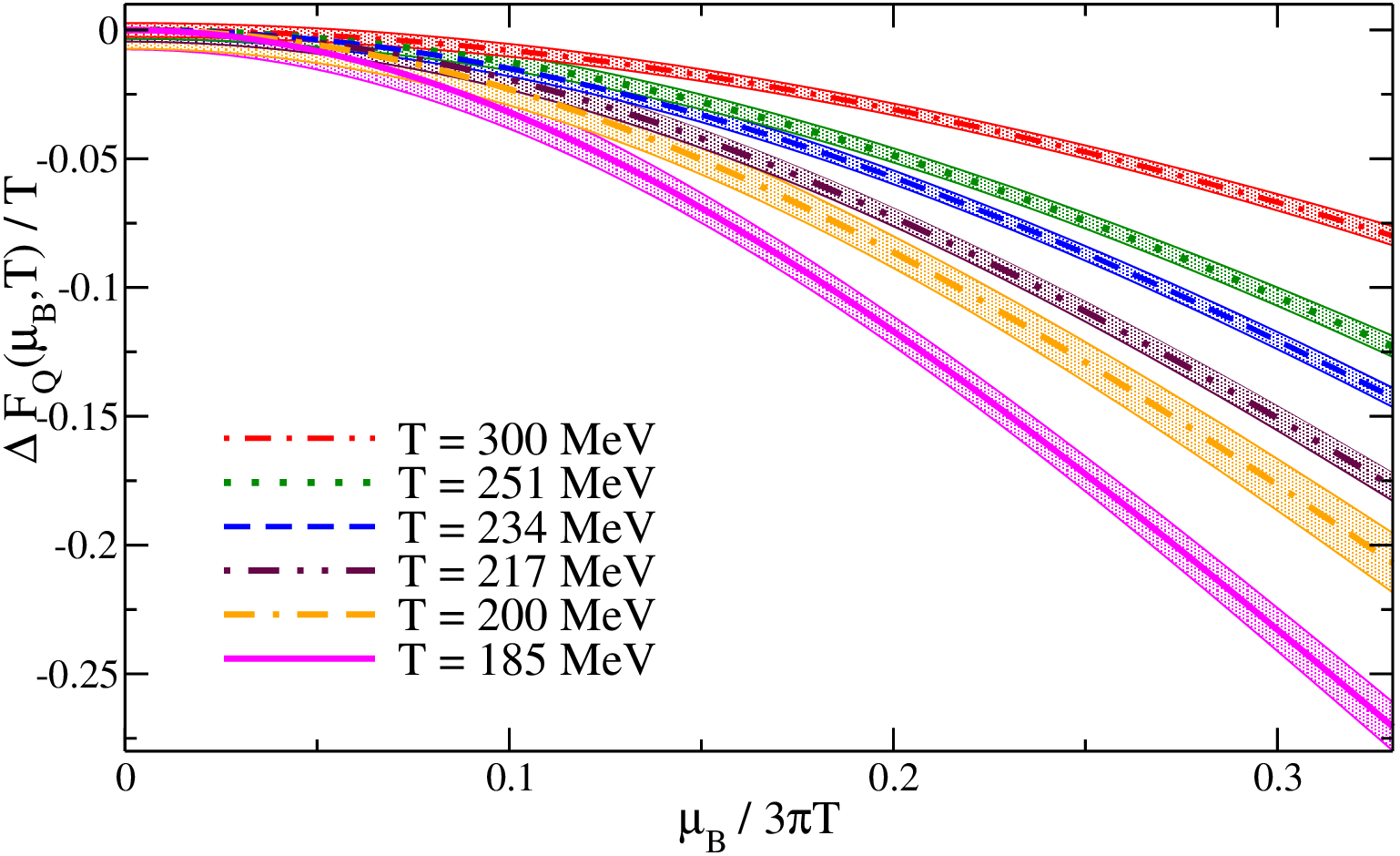}
  \caption{Analytic continuation 
of $\Delta F_Q (\mu_B,T)/T$ to real chemical potentials, obtained 
from a best fit to our numerical data at imaginary
chemical potential and assuming the ansatz
in Eq.~(\ref{ratioloopfit}).}
  \label{fig:fovert_continuation}
\end{figure}

\begin{figure}[t!]
  \includegraphics*[width=\columnwidth]{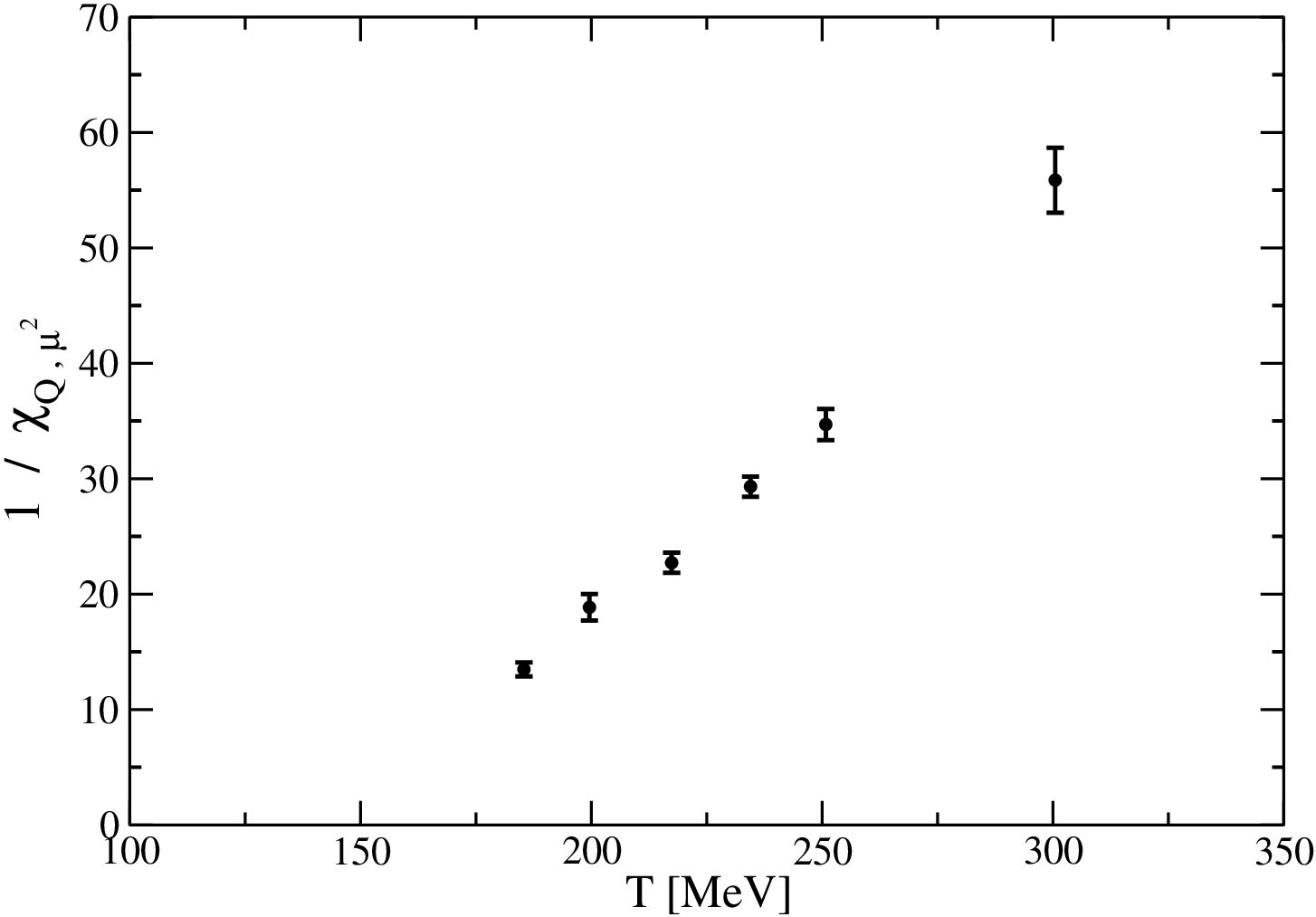}
  \caption{Behavior of the inverse of the coefficient $\chi_{Q,\mu_B^2}$, defined
in Eq.~(\ref{ratioloopfit}), as a function of temperature.}
  \label{fig:fit_invcoeff}
\end{figure}

The analytic continuation of $\Delta F_Q (\mu_B,T)$
to real chemical
potentials according to  Eq.~(\ref{ratioloopfit}) 
is shown in
Fig.~\ref{fig:fovert_continuation}. The free energy of the static quark 
is a decreasing function of $\mu_B$, in agreement with the fact
that a finite baryon chemical potential enhances deconfinement.
Also the fact that the coefficient $\chi_{Q,\mu_B^2}$ 
becomes larger 
as $T$ decreases can be understood qualitatively. Close to the 
pseudocritical temperature, the chemical potential acts as 
a transition driving parameter: if the Polyakov loop were an exact 
order parameter for deconfinement, its dependence on 
$\mu_B$ would become sharper and eventually diverge at the transition.
Of course this is not the case: the Polyakov loop is not 
an order parameter and there is no real transition
in QCD at the physical point; however it is reasonable to expect 
$\chi_{Q,\mu_B^2}$ 
to become larger and larger as $T_c$ is approached from above.

This behavior is 
clearly visible in Fig.~\ref{fig:fit_invcoeff}, where we plot
the inverse coefficient, $1/\chi_{Q,\mu_B^2}$  as a function of $T$.
It is striking to notice that, looking at the high temperature region,
one would be tempted
to predict a vanishing of $1/\chi_{Q,\mu_B^2}$  in a region of temperatures
around 150~MeV, i.e.~roughly coinciding with $T_c$. Of course the behavior
is then smoothed out as $T_c$ is approached more closely.
We cannot assert if this hints at
a more strict connection between
$\Delta F_Q (\mu_B,T)$ and the confinement/deconfinement transition;
however it indicates that this 
is a quantity which is surely worth further investigation.

\begin{figure}[t!]
  \includegraphics*[width=\columnwidth]{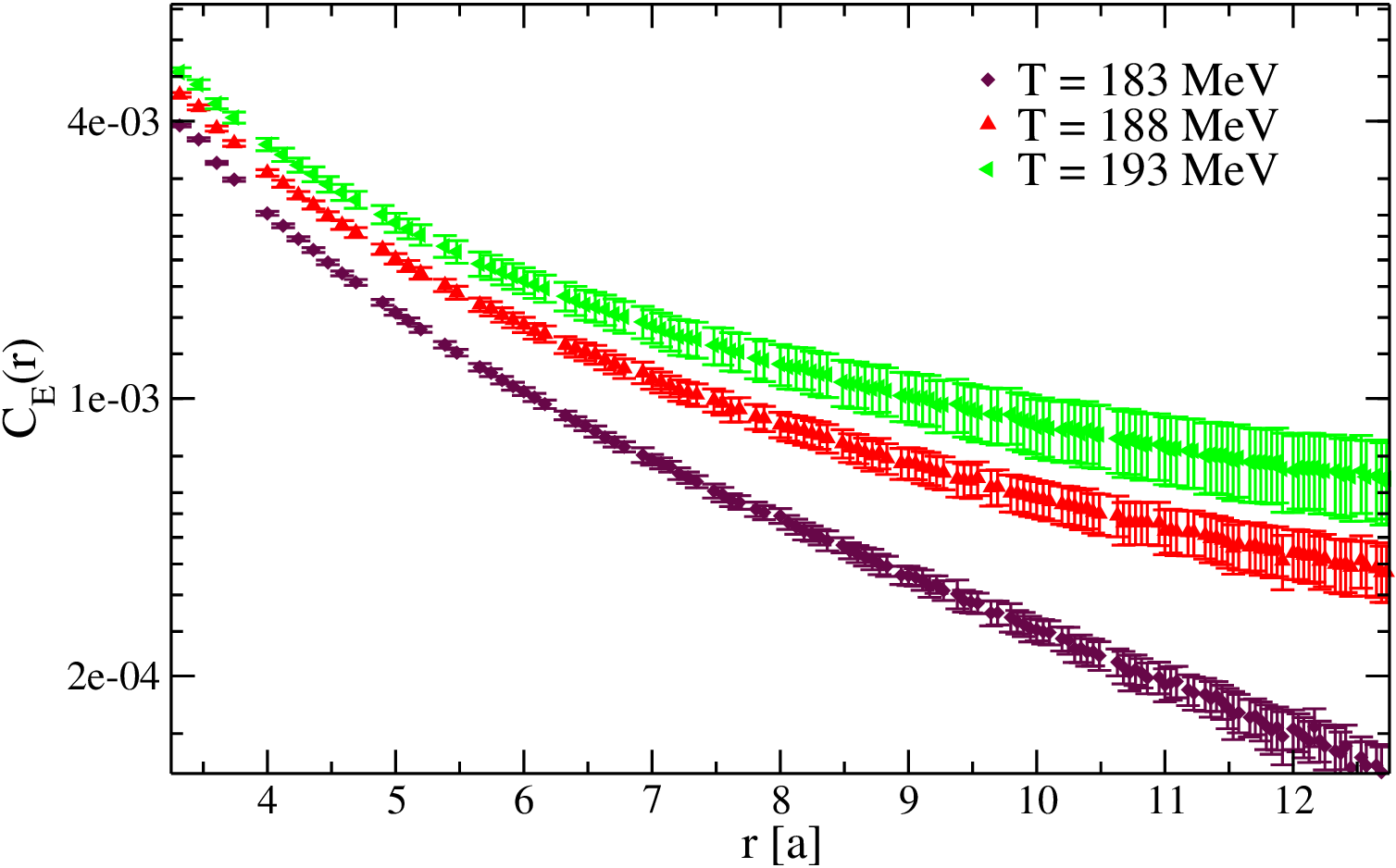}
  \includegraphics*[width=\columnwidth]{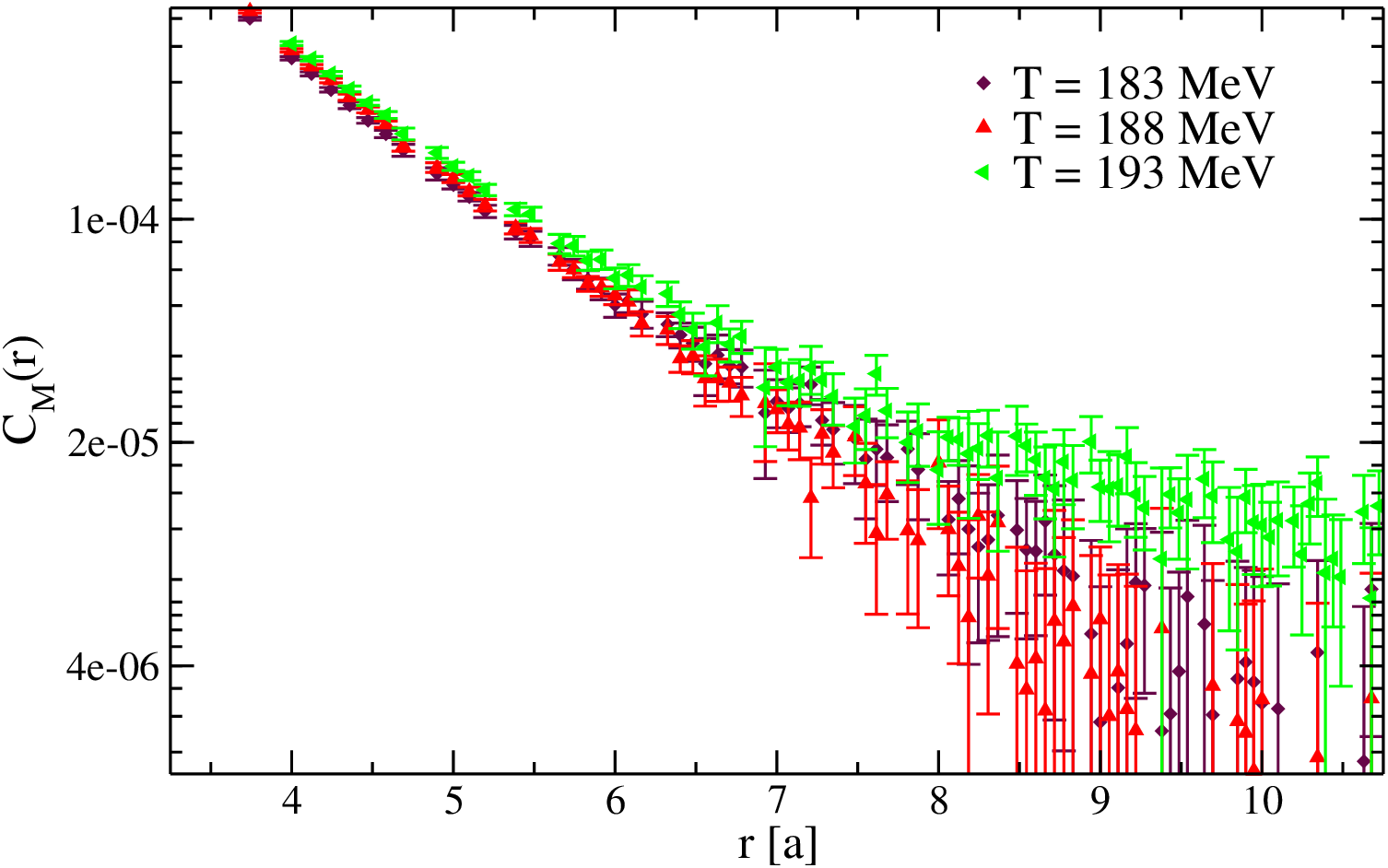}
  \caption{Behavior of the color-electric (top) and color-magnetic
    (bottom) correlators measured on a $40^3 \times 8$ lattice
    at $\mu_I/T = \pi$ and for three different
   temperatures below $T_{RW}(N_t = 8) \sim 200$~MeV.}
  \label{fig:cem_underTc}
\end{figure}

\subsection{Magnetic and electric screening masses close to  the 
Roberge-Weiss endpoint}

As we have stressed above, values of the imaginary chemical potential
which at high $T$ lead to the Roberge-Weiss transition play a special
role. Consider in particular the case $\mu_I/T = \pi$, which corresponds
to a simple shift of fermionic boundary conditions from 
antiperiodic to periodic ones: charge conjugation symmetry is not 
explicitly broken in this case. For $T > T_{RW}$ it gets 
spontaneously broken, so that a mixing between the electric and 
the magnetic sectors appears anyway, however for 
$T < T_{RW}$ it is not and the standard definition of electric
and magnetic correlators, with the associated screening masses,
is well posed, so that it is interesting to compare their 
behavior with the case of zero chemical potential.

\begin{figure}[t!]
  \includegraphics*[width=\columnwidth]{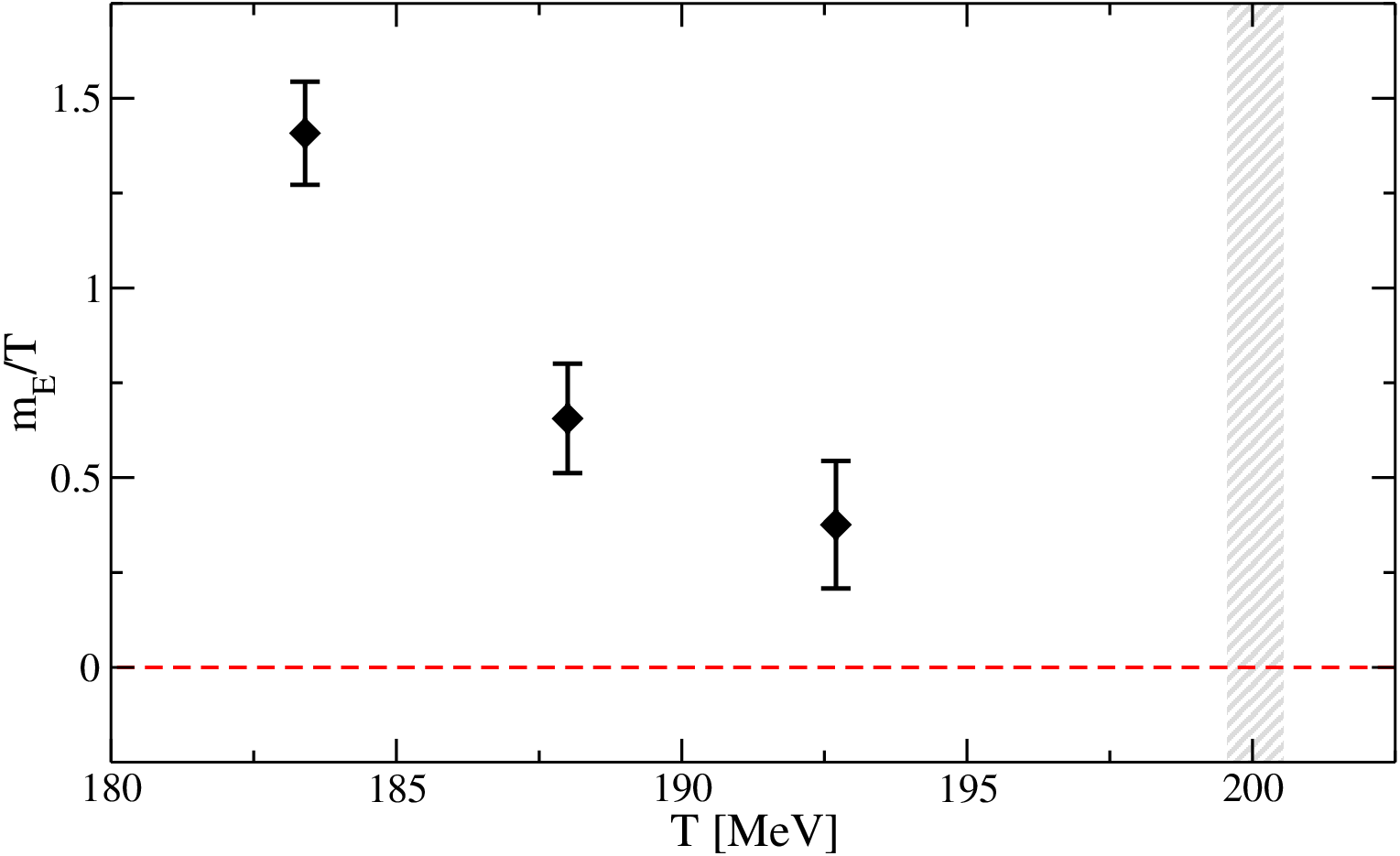}
  \caption{Electric screening masses at $\mu_I/T = \pi$ obtained from
the correlators shown in Fig.~\ref{fig:cem_underTc}. The vertical
gray band indicates the location
of $T_{RW}$ for $N_t = 8$~\cite{noirw}: notice that in this case
we report just the statistical error stemming from 
the determination of the critical coupling in Ref.~\cite{noirw}, and 
not the systematic one stemming from the 
determination of the overall physical scale, which is common
to all data in this limited range of temperatures.}  
  \label{fig:me_underTc}
\end{figure}

In Fig.~\ref{fig:cem_underTc} we show the behavior of such correlators
for three values of $T$. A striking difference with respect to the
$\mu_I/T = 0$ case is clearly visible: the hierarchy is inverted,
with the electric correlators being larger than the magnetic ones
by more than one order of magnitude and growing significantly
as $T$ approaches $T_{RW}$, which for the 
$40^3 \times 8$ lattice that we have used for these measurements
is $T_{RW} (N_t = 8) \sim 200$~MeV~\cite{noirw}.

The reason is easy to understand: the electric correlator
is the two-point function of the imaginary part
of the Polyakov loop (see Eq.~(\ref{eq:mecorrs})), which is also an order
parameter for the Roberge-Weiss transition at $\mu_I/T = \pi$.
Therefore, being directly connected to the order parameter of the transition,
it is expected to undergo the most critical modifications,
due to large distance fluctuations, as one approaches the critical point.

To better clarify this point, we show the values of the electric
screening masses at $\mu_I/T = \pi$ (defined as in 
Eq.~(\ref{eq:mecorrsfit})) as a function of $T$ in
Fig.~\ref{fig:me_underTc}. The electric mass decreases rapidly 
as $T$ approaches $T_{RW}$, seemingly approaching zero 
at the transition, as expected. Notice however 
that, within the present
statistical errors and limited number of data points,
we are not able to check whether we are already close enough 
to $T_{RW}$ to reveal 
the correct critical behavior
predicted for the correlation length at the transition,
which for this case is in the 3D Ising universality class~\cite{noirw}.

\section{Conclusions}
\label{concl}

We have investigated the behavior of Polyakov loops and
of Polyakov loop correlators as a function of the baryon chemical potential
in the deconfined phase of QCD with $N_f = 2+1$ flavors
and physical quark masses.

We have discussed the 
extension of the concept
of gauge invariant screening masses
at finite baryon density, where charge conjugation
symmetry is explicitly broken and a mixing between the electric
and the magnetic sector appears, leading
to the inapplicability of the standard definition of 
color-electric and color-magnetic masses.
Such an extension can be given 
in terms of diagonalized (or alternatively principal) correlators 
derived from a $2 \times 2$ matrix of correlators
involving the real and the imaginary part of the Polyakov loop.
In this way one can obtain a consistent definition 
of two different gauge invariant screening masses
which characterize the thermal medium in the presence 
of a finite baryon density.

In order to obtain a determination of such screening 
masses, we have considered
the theory discretized on $N_t = 8$ lattices by means 
of stout smeared rooted staggered fermions and a tree level 
Symanzik improved gauge action. Numerical simulations have been 
performed for imaginary values of the baryon chemical 
potential, then exploiting analytic continuation.
Both screening masses show an increasing behavior as the 
baryon chemical potential is switched on, and 
the slope of the relative increase (see Eq.~(\ref{eq:massfit})) seems to be 
independent of $T$ and of the same order of 
magnitude as that predicted for the Debye screening mass 
at the lowest order of perturbation theory.
Future investigations should extend present results to 
different values of $N_t$ in order to provide a continuum
extrapolation for the gauge invariant screening masses
at finite baryon density.

We have also shown that for $\mu_I/T = \pi$ and $T < T_{RW}$,
where charge conjugation symmetry is exact and the standard definitions
of electric and magnetic screening masses still holds, the hierarchy
of screening lengths is inverted, with the electric mass being the lowest
one and approaching zero as $T \to T_{RW}$. 
This has been interpreted in terms of the direct
coupling existing between the electric correlator and the 
order parameter for the Roberge-Weiss transition.

Finally, we have investigated the dependence of the static quark free energy
on the chemical potential, which is defined by the large
distance behavior of the Polyakov loop correlators. In particular,
we have considered the free energy variation due to the 
introduction of the chemical potential, $\Delta F_Q (\mu_B,T)$, which
is related to the ratio of Polyakov loops and
is not expected to undergo additive renormalization:
this has been explicitly verified by comparing data
obtained for different values of $N_t$ and different amounts of smearing
on Polyakov loops. The static free energy decreases as a function
of the baryon chemical potential, as expected on general
grounds, with a slope which increases significantly as 
the temperature approaches the pseudocritical temperature
$T_c$ from above. That could hint at a more strict
connection between 
$\Delta F_Q (\mu_B,T)$ and the confinement/deconfinement transition:
future investigations should consider a more systematic 
study of this quantity and its extrapolation
to the continuum limit.

\acknowledgments

We thank A.~Bazavov for useful discussions. 
Numerical simulations have been performed on the GALILEO
machine at CINECA, based on the 
agreement between INFN and CINECA (under project INF17\_npqcd)
and at the Scientific Computing
Center at INFN-PISA. 
FN acknowledges financial support from the INFN HPC\_HTC project.

\end{document}